\documentclass[preprint,prl,preprintnumbers,amsmath,amssymb]{revtex4}
\usepackage{graphicx,epsfig}
\usepackage{dcolumn}
\usepackage{bm}
\begin{document}
\title{Anisotropic translational diffusion in the nematic phase: Dynamical 
signature of the coupling between orientational and translational order in the
energy landscape}
\author{Dwaipayan Chakrabarti and Biman Bagchi} 
\email{bbagchi@sscu.iisc.ernet.in}
\affiliation{Solid State and Structural Chemistry Unit, Indian Institute of 
Science, Bangalore 560012, India}

\begin{abstract}

We find in a model system of thermotropic liquid crystals that the translational 
diffusion coefficient parallel to the director $D_{\parallel}$ first increases and
then decreases as temperature drops through the nematic phase, and this reversal
occurs where {\it the smectic order parameter of the underlying inherent
structures becomes significant for the first time}. We argue, based on an energy
landscape analysis, that the coupling between orientational and translational 
order can play a role in inducing the non-monotonic temperature behavior of 
$D_{\parallel}$. Such a view is likely to form the foundation of a theoretical 
framework to explain the anisotropic translation diffusion.
  
 
\end{abstract}

\maketitle

Anisotropic translational diffusion of non-spherical molecules enjoys immense 
interdisciplinary interests because of its importance in physical 
(liquid crystals), chemical (micelles), and biological (lipids) systems 
\cite{deGennes-book,Kruger-PR,Dvinskikh-JCP-PRE,Johannesson-PRE-1996,
Lettinga-EPL-2005}. It is particularly important in the uniaxial nematic phase, 
where the diffusion description invokes $D_{\parallel}$ and $D_{\perp}$, the 
principal components of the second-rank diffusion tensor, for translational motion
parallel and perpendicular, respectively, to the macroscopic director 
\cite{deGennes-book}. A variety of experimental techniques probe the anisotropic 
translational diffusion in the nematic phase 
\cite{Kruger-PR,Dvinskikh-JCP-PRE,Lettinga-EPL-2005}. 
However, a consensus regarding an appropriate dynamical model still lacks. In
particular, the role of coupling between the orientational and translational order
parameters appears to be overlooked.

On the contrary, the interplay between orientational and translational order has 
been extensively discussed in the context of the nematic-smectic-A (NA) phase 
transition over three decades \cite{deGennes-book,McMillan-PRA,deGennes-SSC,
HLM-PRL-SSC,Dasgupta-PRL,Cladis-PRL-1989,Anisimov-PRA-1990,Yethiraj-PRL-2000,
Lelidis-PRL}. The de Gennes-McMillan (dGM) coupling, which refers to the 
occurrence of the smectic (one-dimensional translational) 
ordering being intrinsically coupled with increase in the nematic (orientational) 
ordering \cite{deGennes-book,McMillan-PRA,deGennes-SSC}, could drive, within a 
mean field approximation, an otherwise continuous NA transition first
order for a narrow nematic range \cite{McMillan-PRA}. Halperin, Lubensky, and
Ma later invoked the coupling between the smectic order parameter and the
transverse director fluctuations in their theoretical treatment that predicted NA
transition to be {\it at least weakly first order} \cite{HLM-PRL-SSC}.
 
Intuitively, $D_{\parallel}$ appears to be well placed to capture the dynamical
signature of the coupling between orientational and translational order. 
Therefore, we here investigate anisotropic translational diffusion in a model 
system of thermotropic liquid crystals. The observed diffusion behavior of the 
system is correlated with the features of its underlying potential energy 
landscape \cite{Wales-book}. In this Letter, we show that the coupling between 
orientational and translational order can lead to the {\it non-monotonic 
temperature behavior} of $D_{\parallel}$ being reported here.  

We have investigated a system of $256$ ellipsoids of revolution along two 
isochors at a series of temperatures. We have used the well-established Gay-Berne 
(GB) pair potential \cite{Gay-Berne}, which explicitly incorporates anisotropy in 
both the attractive and the repulsive parts of the interaction with a single-site 
representation for each ellipsoid of revolution \cite{Brown-PRE}. The GB pair 
potential gives rise to a family of models, each member of which is characterized 
by the values chosen for the set of four parameters
$(\kappa, \kappa^{\prime}, \mu, \nu)$ \cite{Miguel-JCP}. Here $\kappa$ defines the
aspect ratio, that is the ratio of molecular length to breadth, of the ellipsoid 
of revolution and $\kappa^{\prime}$ is the energy anisotropy parameter defined by 
the ratio of the depth of the minimum of the potential for a pair of molecules 
aligned parallel in a side-by-side configuration to that in an end-to-end 
configuration while $\mu$ and $\nu$ are two adjustable exponents that also control
the anisotropy in the well depth \cite{Miguel-JCP}. We have employed
the original and most studied parameterization: $\kappa = 3, \kappa^{\prime} = 5, 
\mu = 2, \nu = 1$ \cite{Gay-Berne,Miguel-JCP}. The isochors have been so chosen
that the range of the nematic phase along these varies considerably. 

\begin{figure}
\epsfig{file=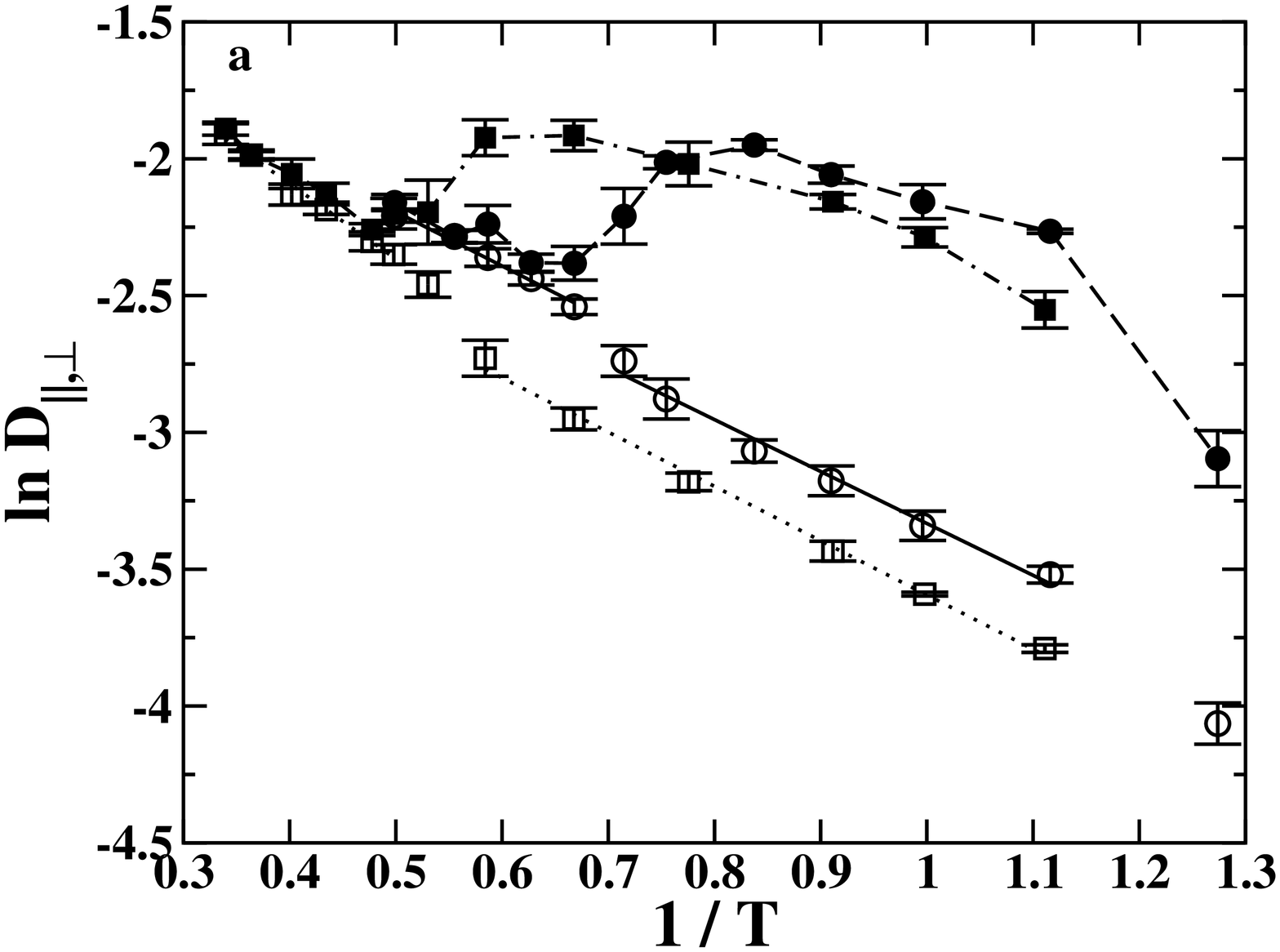,width=5.5cm,angle=270}
\epsfig{file=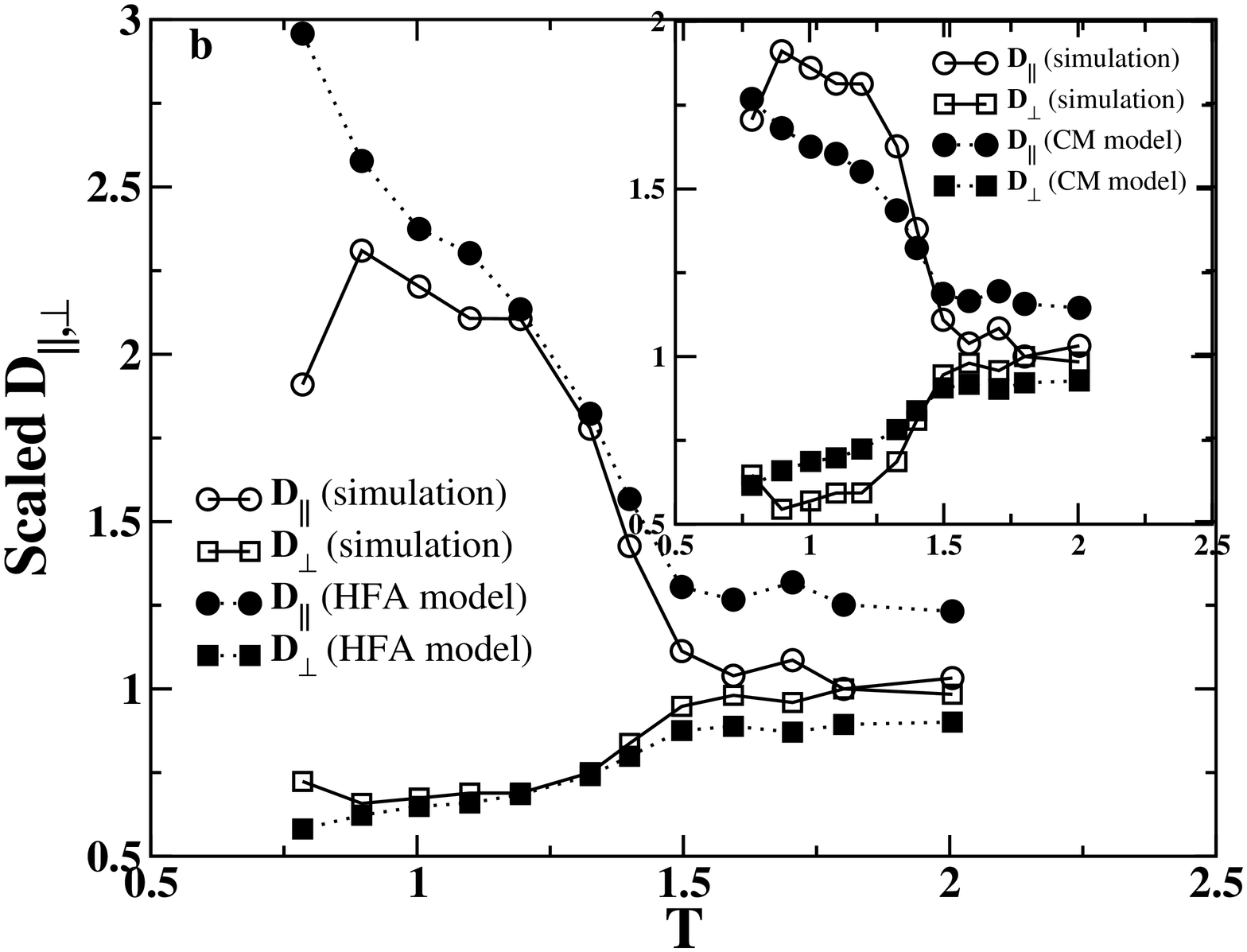,width=5.5cm,angle=270}
\caption{(a) The self-diffusion coefficients $D_{\parallel}$ and $D_{\perp}$  
in the logarithmic scale versus the inverse temperature along two isochors at
densities $\rho = 0.32$ (circles) and $0.33$ (squares), respectively. The 
dot-dashed and long-dashed lines lines are guide to eye for the $D_{\parallel}$ 
data (filled symbols) and the solid lines and the dotted lines are the Arrhenius 
fits to the $D_{\perp}$ (empty symbols) data for $\rho = 0.32$ and $0.33$, 
respectively. $D_{\perp}$ data have been considered separately across the 
isotropic phase and the nematic phase for the Arrhenius fits. (b) The comparison 
of the scaled $D_{\parallel}$ and $D_{\perp}$ data obtained from our simulations 
with those predicted by the Hess-Frenkel-Allen (HFA) model (main frame) and the 
Chu and Moroi (CM) model (inset). For the comparison with the HFA model the 
scaling is done by $\langle D \rangle _{g}$ while for that with the CM model it is
done by $\langle D \rangle$.}
\end{figure}
Fig. 1a shows the inverse temperature dependence of the principal components of 
the diffusion tensor (in the logarithmic scale) of the Gay-Berne system with the 
aspect ratio 3 along the two isochors considered \cite{method}. $D_{\parallel}$ 
and $D_{\perp}$ are obtained from the slopes at long times of the respective mean 
square displacements versus time plots: 
$D_{\parallel} = \frac{1}{2} \lim_{t \rightarrow \infty} 
\frac{d}{dt}<\Delta r_{\parallel}^{2}(t)>, D_{\perp} = \frac{1}{2} 
\lim_{t \rightarrow \infty} \frac{d}{dt}<\Delta r_{\perp}^{2}(t)>$,
where
$<\Delta r_{\parallel}^{2}(t)> = <(r_{\parallel}(t)-r_{\parallel}(0))^2>$ and
$<\Delta r_{\perp}^{2}(t)> = <(r_{\perp}(t)-r_{\perp}(0))^2>$ 
\cite{Bates-Luckhurst-JCP-2004}. Here the subscripts refer to the Cartesian 
components, resolved in a system of axes based on the director defined at each 
time origin. For the finite size of the system, the average orientational order 
parameter $S$ has a nonzero value even in the isotropic phase 
\cite{order-parameter}. This allows us to compute $D_{\parallel}$ and $D_{\perp}$ 
also in the isotropic phase. It is evident from Fig. 1a that both $D_{\parallel}$ 
and $D_{\perp}$, which have nearly identical values in the isotropic phase, 
exhibit an Arrhenius temperature dependence in this phase. On crossing the 
isotropic-nematic (I-N) phase boundary as temperature drops, $D_{\parallel}$ {\it 
first increases and then decreases} while $D_{\perp}$ continues to undergo a 
monotonic decrease following an Arrhenius temperature behavior across the nematic 
phase. From the Arrhenius fits to the $D_{\perp}$ data, we find that the 
activation energy for the diffusive translational motion perpendicular to the 
director remains effectively unchanged on either side of the I-N transition.

A quantitative, {\it albeit indirect}, approach to capture the dynamical signature 
of the coupling between orientational and translational order is to compare the 
$D_{\parallel}$ and $D_{\perp}$ data obtained from our simulations with those 
predicted by the existing dynamical models, which ignore such coupling. In 
Fig. 1b, we do so by considering two theoretical models \cite{CM,HFA}, that have 
been applied to trace experimental and molecular dynamics simulation data of 
anisotropic translational diffusion in the nematic phase of liquid crystalline 
systems \cite{CM,HFA,Leadbetter-CPL,Urbach-JCP,Dvinskikh-JCP-PRE}. The main frame 
displays the comparison with the Hess-Frenkel-Allen (HFA) model while the inset 
shows the same with the Chu and Moroi model. The latter gives relatively simple 
expressions for $D_{\parallel}$ and $D_{\perp}$ in terms of {\it only} the 
orientational order parameter $S$ and the shape factor $g = \pi / (4 \kappa)$:
$D_{\parallel} = \langle D \rangle [1 + 2 S (1 - g) / (2g + 1)]$ and 
$D_{\perp} = \langle D \rangle [1 - S (1 - g) / (2g + 1)]$,
where the isotropic average is defined by $\langle D \rangle = (2D_{\perp} +
D_{\parallel}) / 3$. The HFA model invokes the concept of affine transformation 
from the space of isotropic hard spheres and yields the following expressions:
$D_{\parallel} = \langle D \rangle _{g} \alpha [\kappa^{4/3} - 2/3 \kappa^{-2/3}
(\kappa^{2} - 1) (1 - S)]$ and
$D_{\perp} =  \langle D \rangle _{g} \alpha [\kappa^{-2/3} + 1/3 \kappa^{-2/3}
(\kappa^{2} - 1) (1 - S)]$, where
$\alpha = [1 + 2/3 (\kappa^{-2} - 1) (1 - S)]^{-1/3} [1 + 1/3 (\kappa^{2} - 1) 
(1 - S)]^{-2/3}$
and the geometric average is defined by $\langle D \rangle_{g} = D_{\perp}^{2/3}
D_{\parallel}^{1/3}$. For the purpose of comparison, we plot scaled 
$D_{\parallel}$ and $D_{\perp}$ data. It follows from Fig. 1b that both the models
cannot capture the non-monotonic temperature behavior of
$D_{\parallel}$. We next demonstrate {\it directly} by performing a landscape
analysis that the non-monotonic temperature behavior of $D_{\parallel}$ could be 
due to the the coupling between orientational and translational order.

\begin{figure}
\epsfig{file=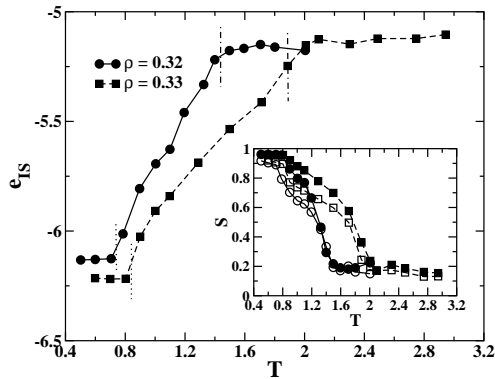,width=5.5cm,angle=270}
\caption{The temperature dependence of the average inherent structure energy 
per particle along the two isochors at densities $\rho = 0.32, 0.33$. The inset 
shows the evolution of the average orientational order parameter S with 
temperature both for the inherent structures (filled) and the corresponding 
pre-quenched ones (empty). The two sets of data are for the same two densities as 
in Fig. 1a. The vertical dot-dashed and dotted lines in the main frame show
the locations of the isotropic-nematic and nematic-smectic phase boundaries, 
respectively.}
\end{figure}
In the landscape formalism, the potential energy surface is partitioned into a 
large number of "basins", each defined as the set of points in the 
multidimensional configuration space such that a local minimization of the 
potential energy maps each of these points to the same local minimum 
\cite{Wales-book}. The inherent structure corresponds to the minimum configuration
\cite{Stillinger-PRA}. As a result of this partitioning of the configuration 
space, the time dependent position ${\bf r}_{i}(t)$ of a particle $i$ can be 
resolved into two components: ${\bf r}_{i}(t) = {\bf R}_{i}(t) + {\bf S}_{i}(t)$, 
where ${\bf R}_{i}(t)$ is the spatial position of the particle $i$ in the inherent
structure for the basin inhabited at time $t$, and ${\bf S}_{i}(t)$ is the 
intrabasin displacement away from that inherent structure \cite{Shell-JPCB}. That 
the replacement of the real positions ${\bf r}_{i}(t)$ by the corresponding 
inherent structure positions in the Einstein relations yields an equivalent 
diffusion description, as has been theoretically argued and also verified in 
simulations \cite{Shell-JPCB,Keyes-PRE}, is the key to our analysis presented 
here.
 
Fig. 2 displays the average inherent structure energy as the drop in temperature 
drives the system across the mesophases along two different isochors 
\cite{pel-method,Chakrabarti-preprint}. In the inset of Fig. 2, we show the 
concomitant evolution of the average orientational order parameter $S$ both for 
the inherent structures and the corresponding pre-quenched ones. While the average 
inherent structure energy remains fairly insensitive to temperature variation in 
the isotropic phase and also in the smectic phase, it undergoes a steady fall as 
the orientational order grows through the nematic phase. We find that 
$D_{\parallel}$ starts increasing near the I-N phase boundary at a temperature 
that {\it marks the onset of the growth of the depth of the potential energy 
minima explored by the system}.

\begin{figure}
\epsfig{file=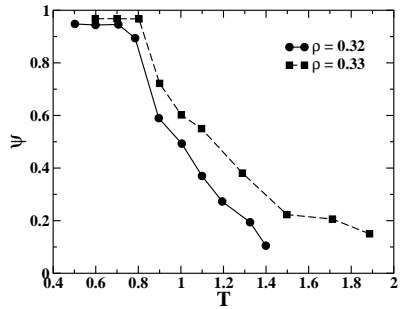,width=5cm,angle=270}
\caption{The evolution of the smectic order parameter $\Psi$ for the inherent 
structures with temperature at two densities.}
\end{figure}
The onset of the growth of the orientational order in the vicinity of the I-N 
transition {\it induces a translational order in a layer in the underlying 
quenched configurations} \cite{Chakrabarti-preprint}. The smectic
order parameter $\Psi$ provides a quantitative measure of the one-dimensional 
translational order \cite{sop}. In Fig. 3, we show the evolution of the average
smectic order parameter $\Psi$ of the inherent structures, obtained by averaging 
over the quenched configurations, with temperature \cite{sop}. {\it A steady 
increase in $\Psi$ with the concomitant growth of $S$ in the underlying inherent 
structures is apparent across the nematic phase.} 

\begin{figure}
\epsfig{file=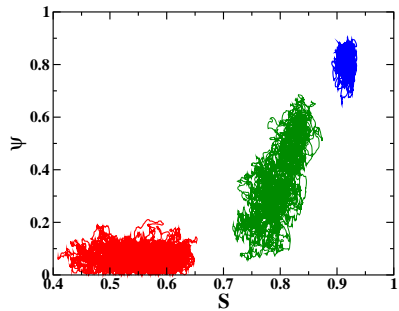,width=5cm,angle=270}
\caption{The coupling between the orientational order and translational order in 
the Gay-Berne system with the aspect ratio $\kappa= 3$ at three state points 
along the isochor at density $\rho = 0.32$: at the nematic phase 
($T = 1.194$; red), at the smectic phase ($T = 0.502$; blue), and at the 
nematic-smectic transition region ($T = 0.785$; green). Here $S$ and $\Psi$ denote
the respective order parameters for instantaneous configurations.}
\end{figure}
The interplay between the orientational order and translational order, shown in 
Fig. 3, is reminiscent of the dGM coupling which was originally 
conceived to be present near the nematic-smectic phase boundary in the parent 
system. Figure 4 confirms this with an explicit demonstration of the coupling 
between the smectic order parameter $\Psi$ and the nematic order parameter $S$ 
near the nematic-smectic transition region. While the fluctuation of $S$ is large 
at the nematic phase, it is the fluctuation of $\Psi$ that is rather large in the
smectic phase. A strong coupling between the two is evident at the nematic-smectic
transition region where configurations with larger $S$ values tend to have larger 
$\Psi$ values. 

On scrutiny of Figs. 1 and 3, we find that the reversal in the temperature 
behavior of $D_{\parallel}$ in the nematic phase as temperature drops occurs when 
the the smectic order parameter in the underlying inherent structures becomes 
significant (above 0.3) for the first time. The smectic order parameter is a
measure of the translational order which appears in a layer perpendicular to
the director. The induction of such translational order makes translational motion
parallel to the director much difficult, resulting in a reducing effect on
$D_{\parallel}$. From the viewpoint of the energy landscape analysis, 
translational order in the underlying inherent structures therefore appears to 
play a key role in the non-monotonic temperature dependence of $D_{\parallel}$. 
The latter can therefore be taken as a {\it dynamical signature of the de 
Gennes-McMillan coupling augmented in the potential energy landscape}. 

System size in the present study has been optimal given that the landscape 
studies have often been restricted to a smaller system size while long wavelength 
fluctuations in the vicinity of a phase transition suggest a bigger one to be 
undertaken. The system size we have chosen here is, however, large enough so that 
the system tracks the phase diagram reported earlier \cite{Miguel-JCP}. 
Nevertheless, in order to check possible system size effects on our results, we 
have further considered systems with $500$ ellipsoids of revolution along the 
isochor at density $\rho = 0.32$. {\it No qualitative change} in the results has 
been observed (data not shown). 

We have further studied effects of varying the aspect ratio $\kappa$ and the 
energy anisotropy parameter $\kappa^{\prime}$ separately to explore the robustness
of our results and analysis. In particular, we have considered the aspect ratio 
$\kappa = 3.8$ along the isochor at $\rho = 0.235$, for which a stable smectic-A 
phase appears between a wide nematic and low-temperature smectic-B phase. The 
temperature behavior of $D_{\parallel}$ in the nematic phase has been found to be 
qualitatively similar to what has been observed with the aspect ratio $\kappa = 
3$, for which the smectic-A phase appears only in the underlying inherent 
structure \cite{Chakrabarti-preprint}. It is particularly interesting to consider 
a case where the smectic phase is absent and contrast the behavior. To this end, 
we have considered $\kappa^{\prime} = 1$, for which no smectic phase appears even 
at low temperatures and the underlying inherent structures for the nematic phase 
also do not have on the average any translational order 
\cite{Chakrabarti-preprint}. In this case, we find that the signature of the
non-monotonic behavior in the temperature dependence of $D_{\parallel}$ is rather
weaker and is even missing in the scaled data (data not shown) -- the dynamical 
models considered here also provide a better description of the anisotropic 
translational diffusion data. This further substantiates the importance of the
coupling between orientational and translational order in the anisotropic 
translation diffusion in the nematic phase. 

In summary, the present work throws light on the plausible role of the coupling 
between orientational and translational order in inducing a non-monotonic 
temperature behavior of $D_{\parallel}$ in the nematic phase. While the 
competition between the alignment and thermal effects can also give rise to a 
non-monotonic behavior, the importance of such coupling cannot be ignored 
particularly when a low-temperature smectic phase exists. A comparison of the 
simulated $D_{\parallel}$ data with those predicted by two well-known theoretical 
models shows the inadequacy of these models to capture the observed non-monotonic 
temperature dependence of $D_{\parallel}$. The energy landscape analysis presented
here suggests the necessity of a theoretical treatment that includes the coupling 
between orientational and translational order, which has an augmented 
manifestation in the underlying energy landscape. Such a suggestion is likely to 
form the foundation of a theoretical framework to explain the features anisotropic
translational diffusion.

We thank C. Dasgupta and S. Ramaswamy for useful discussions. This work was 
supported in parts by grants from DST and CSIR, India. DC acknowledges UGC, India 
for providing financial support.


\begin{references}
\bibitem{method} Molecular dynamics simulations have been performed in a simple 
cubic box with periodic boundary conditions. All quantities are given in reduced 
units defined in terms of the Gay-Berne potential parameters $\epsilon_{0}$ and 
$\sigma_{0}$: length in units of $\sigma_{0}$, temperature in units of 
$\epsilon_{0}/k_{B}$, $k_{B}$ being the Boltzmann constant, and time in units of 
$(\sigma_{0}^{2}m/\epsilon_{0})^{1/2}$, $m$ being the mass of the ellipsoids of 
revolution. We have set the mass as well as the moment of inertia of the 
ellipsoids equal to unity. See Ref. \cite{Brown-PRE} for implication of the 
symbols. The intermolecular potential has been truncated at a distance and shifted
as in Ref. \cite{Brown-PRE}. Equilibration has been done by periodic rescaling of 
linear and angular velocities of particles while the data collection has been in a microcanonical ensemble.
\end{references}
\end{document}